# Title: Electronic spin transport in dual-gated bilayer graphene

Ahmet Avsar[1*], Ivan Jesus Vera-Marun[2], Jun You Tan[1], Gavin Kok Wai Koon[1], Kenji Watanabe[3], Takashi Taniguchi[3], Shaffique Adam[1] and Barbaros Özyilmaz[1*]

**The elimination of extrinsic sources of spin relaxation is key in realizing the exceptional intrinsic spin transport performance of graphene. Towards this, we study charge and spin transport in bilayer graphene-based spin valve devices fabricated in a new device architecture which allows us to make a comparative study by separately investigating the roles of substrate and polymer residues on spin relaxation. First, the comparison between spin valves fabricated on $SiO_2$ and BN substrates suggests that substrate-related charged impurities, phonons and roughness do not limit the spin transport in current devices. Next, the observation of a 5-fold enhancement in spin relaxation time in the encapsulated device highlights the significance of polymer residues on spin relaxation. We observe a spin relaxation length of ~ 10 μm in the encapsulated bilayer with a charge mobility of 24000 $cm^2$/Vs. The carrier density dependence of spin relaxation time has two distinct regimes; $n < 4 \times 10^{12}$ $cm^{-2}$, where spin relaxation time decreases monotonically as carrier concentration increases, and $n \geq 4 \times 10^{12}$ $cm^{-2}$, where spin relaxation time exhibits a sudden increase. The sudden increase in the spin relaxation time with no corresponding signature in the charge transport suggests the presence of a magnetic resonance close to the charge neutrality point. We also demonstrate, for the first time, spin transport across bipolar p-n junctions in our dual-gated device architecture that fully integrates a sequence of encapsulated regions in its design. At low temperatures, strong suppression of the spin signal was observed while a transport gap was induced, which is interpreted as a novel manifestation of impedance mismatch within the spin channel.**

[1]Centre for Advanced 2D Materials & Department of Physics, 2 Science Drive 3, National University of Singapore, Singapore 117542, Singapore; [2]The School of Physics and Astronomy, The University of Manchester, M139PL, Manchester, United Kingdom; [3]National Institute for Materials Science, 1-1 Namiki, Tsukuba 305-0044, Japan.

Correspondence:

Prof. B Özyilmaz, 2 Science Drive 3, S13-02-10, Department of Physics, National University of Singapore, 117542, Singapore. Tel: (65) 6516 6979. E-mail: barbaros@nus.edu.sg

or Dr. A Avsar, 2 Science Drive 3, S13-02-13, Department of Physics, National University of Singapore, 117542, Singapore. Tel: (65) 6516 2532. E-mail: c2davsa@nus.edu.sg

# INTRODUCTION

Graphene is considered to be a promising spin channel material for future spintronics applications[1] due to its high electronic mobility[2], weak spin orbit coupling[3,4] and negligible hyperfine interaction[5,6]. The initial spin transport studies were mainly performed in single layer[7–10] and bilayer exfoliated graphene[9,11] and large area graphene[12–15] deposited on conventional $SiO_2$ substrates. While enhanced spin relaxation times were reported for bilayer graphene-based devices compared to those in single layer, the relatively low spin diffusion constants overall yielded lower spin relaxation length of only 1-2 μm[9,11], far below the theoretical predictions[16]. One approach suggested for achieving longer distance spin communication is to increase the spin diffusion constants by fabricating higher mobility devices[8,17].

For charge transport, it has already been shown that the carrier mobility of graphene devices on $SiO_2$ is mainly limited by interfacial charged impurities, surface roughness and phonons[18–20]. The demonstration of an order-of-magnitude improvement in the mobility of graphene encapsulated between atomically flat, charge trap free boron nitride crystals[21,22] has triggered the recent spin transport studies in encapsulated single layer and recently bilayer graphene-based spin valves where spin relaxation lengths up to ~ 12 μm and ~ 24 μm have been observed, respectively[23,24]. For the case of bilayer graphene, the initial detailed experiments on $SiO_2$ revealed an inverse scaling between spin and momentum relaxation times, e.g. the longest spin relaxation times were observed in the lowest mobility devices[9,11]. Within the standard picture of spin relaxation mechanism, such scaling was interpreted as the dominance of Dyakonov-Perel type scattering mechanism[25] in bilayer graphene-based spin valves on $SiO_2$. Nevertheless, the strength of the extracted spin-orbit coupling was one order of magnitude higher than the theoretical predictions which brings into question this standard interpretation[26]. Recently, Kochan *et al.* proposed a new spin relaxation mechanism beyond the standard picture for both single layer and bilayer graphene driven by resonant scattering due to low concentration of magnetic impurities[27,28]. For the testing of this theory, bilayer graphene is a more ideal platform since a nonmonotonic correlation is predicted for the energy dependence of the spin relaxation time in bilayer graphene case that can be experimentally verified[27]. Ultimately similar mobility enhancement in bilayer graphene may be a successful route to enhance the spin relaxation lengths, although this is strongly dependent on the scattering source and the mechanism of spin relaxation[27,29–31].

## EXPERIMENTAL PROCEDURE

**Preparation of the devices**

In this letter, we discuss the charge and spin transport in high mobility bilayer graphene-based spin valve devices, using a novel device architecture that overcomes the current technological limit of having a single encapsulated region[23,24]. In order to understand the effect of substrate and polymer residues on spin relaxation, we have fabricated three types of bilayer graphene spin valve devices in a single sample; 1- ) spin valve on SiO$_2$, 2- ) spin valve on BN, 3- ) encapsulated spin valve between BN substrate and pre-patterned BN strip. The schematics for the fabrication of encapsulated spin valve devices are shown in Figure 1 a. The fabrication starts with the transfer of a bilayer graphene onto atomically flat BN crystals by utilizing the standard dry transfer method[32]. In order to create uniform and ultra clean interfaces, graphene on BN substrate is first etched with O$_2$ plasma into 1 μm width strip in the bubble and wrinkle free regions of the stack and then annealed at 340 °C for 6 hours under Ar/H$_2$ gaseous mixture (Figure 1b). This annealing process minimizes the fabrication residues formed during the etching processes. Meanwhile, a second BN layer is exfoliated onto SiO$_2$ (300 nm)/Si wafer, then patterned into BN strips by using e-beam lithography and etching (CHF$_3$-O$_2$-, 10-1 ratio) processes (Figure 1c). This pre-patterned BN layer, a novel aspect in our device architecture, is isolated from the wafer with a KOH etching process and finally transferred onto previously prepared BLG/BN stack as shown in Figure 1d. The final stack is annealed under the same conditions as above. The fabrication is completed by forming MgO / Co / Ti (2.2 nm / 30 nm / 5 nm) electrodes on top of both the h-BN strips and non-encapsulated regions of the graphene strip (Figure 1e). These serve as top gate electrodes and direct contact electrodes to bilayer graphene, respectively. The width of the contact electrodes was varied from 300 nm to 1 μm in order to ensure the different coercive fields to switch the relative polarization directions of the ferromagnetic electrodes during spin transport measurements. The distance between edges of adjacent contacts is ~ 3 μm and the width of encapsulating BN strip is 2.5 μm. The details for the fabrication of the MgO barrier layer are discussed in the Supplementary Information S1. Even though the annealing process after metallization is a standard procedure to achieve very high mobilities in graphene-based heterostructure devices[18,19], we omitted this process due to the degradation of ferromagnetic electrodes during annealing (Supplementary Figure S5)[8]. The pre-patterned top BN layer therefore adds a new degree of freedom to the device architecture, as it allows to easily design a series of encapsulated regions with arbitrary lengths, whereas the non-encapsulated regions destined for the contacts can be scaled down to the lower precision limit of the lithography technique. This represents a technological improvement of the prevalent approach in literature, limited in practice to a single encapsulated region[23,24].

**Electrical measurements of the devices**

Measurements are carried out with standard ac lock-in techniques at low frequencies using the local four-terminal setup for charge transport measurements and the nonlocal setup for spin transport measurements (Figure 1 a). In the local charge transport measurements, the current flows between electrode *1* and electrode *4* and a local voltage drop is measured between electrode *2* and electrode *3*. In the non-local measurement configuration, the current flows between the pair of electrodes *1* (injector) and *2*, and a non-local voltage is recorded across the neighboring pair of electrodes *3* (detector) and *4*. SiO$_2$ and h-BN dielectrics are utilized to apply back and top gate biases ($V_{BG}$ and $V_{TG}$), respectively. Unless otherwise stated, electrode *6* is used to locally tune the carrier type and concentration (Supplementary Figure S4). All transport measurements are performed under vacuum conditions (~ 1 x 10$^{-6}$ Torr). In this work, we have characterized a total of 5 junctions in two separate samples (Samples A and B). The bilayer graphene in Sample A has both encapsulated and non-encapsulated junctions, allowing for a direct comparison of charge and spin transport for both cases (Supplementary Figure S1). Optical image of the completed Sample B is shown in Figure 1e.

## RESULTS AND DISCUSSIONS

We first discuss the room temperature spin transport results of the non-encapsulated device in Sample A at $V_{BG}$ = 0 V. The relative magnetization directions of the Co electrodes are altered by sweeping the in-plane magnetic field along their easy axis. This changes the spin accumulation between the injector and detector electrodes, and hence leads to a nonlocal spin signal with a change in resistance $\Delta R$ of ≈ 0.3 Ω (Figure 2a). We note that the observed spin signal is comparable to the earlier reports[11,12,33] (Supplementary Figure S6). In order to confirm the origin of the spin signal and extract the important spin parameters, we performed Hanle spin precession measurements. For this purpose, the magnetization directions of the injector and detector electrodes are made parallel to each other by the application of an in-plane magnetic field. This is followed by sweeping the magnetic field perpendicular to the graphene plane forcing the spins to precess (Figure 2b). The obtained spin precession data is fitted with the solution of the Bloch equation, $R_{NL} \sim \int_0^\infty \frac{1}{\sqrt{4\pi D_S t}} exp\left(\frac{-L^2}{4D_S t}\right) exp\left(\frac{-t}{\tau_S}\right) cos(w_L t) dt$, where $L$ (≈ 1.7 μm) is the separation between the electrodes (center-to-center distance) and $\omega_L$ is the Larmor frequency. This gives a spin relaxation time ($\tau_S$) of ≈ 87 ps, a spin diffusion constant ($D_S$) of ≈ 0.02 m²/s, and hence, a spin relaxation length ($\lambda_S = \sqrt{D_S \tau_S}$) of ≈ 1.3 μm. Next, we characterized the adjacent, encapsulated device in the same sample. Clear spin valve and Hanle spin precession signals are obtained in this device as well (Figure 2-b&c). For this junction, we extracted a spin relaxation time $\tau_S$ ≈ 416 ps, a spin diffusion constant of $D_S$ ≈ 0.089 m² / s, and, hence, a spin relaxation length of $\lambda_S$ ≈ 6.1 μm using the same fitting procedure as for the non-encapsulated junction. These measurements indicate four important conclusions. 1-) The extracted spin parameters in our non-encapsulated bilayer graphene are comparable with those extracted for bilayer devices on SiO$_2$ with similar mobilities[11] (Inset of Figure 2d and See Supplementary Figure S2). This suggests that substrate-related issues such as roughness, interfacial charged impurities and surface phonons are not the limiting source of spin relaxation in current devices. 2- ) Since the encapsulated region is protected against polymer residues during the contact fabrication step, the 5-fold enhancement of spin parameters in the encapsulated device compared to the non-encapsulated one suggests that residues have a significant effect on spin transport. 3- ) The observation of enhancement in the encapsulated device compared to the non-encapsulated one suggests that the contacts (~ 5 kΩ) are not the limiting factor for the spin relaxation. Note that the obtained spin parameters for the encapsulated device should be considered as lower bounds, since they include the spin transport in the non-encapsulated contact regions. Our observation now allows us to speculate that the observed long spin life times in ref. [34] could be attributed mainly to the polymer free fabrication of graphene spin valves rather than only the improvement in contacts. This is also consistent with the observation of nearly 3-fold enhancement of up to 2.7 ns spin relaxation time in single layer graphene upon plasma hydrogenation treatment, even when supported on $SiO_2$, which could be partially attributed to the removal of such residues[35].

Now we discuss the origin of the spin scattering mechanism in bilayer graphene. Figure 2d shows the mobility dependence of spin relaxation time where we observe an increase in spin relaxation time as the mobility increases. The increase of mobility is mainly achieved with decreasing the concentration of the impurity. The latter indicates that impurity scattering has a significant role in spin scattering, contrary to previous studies that have demonstrated that while charged impurities affect the carrier mobility they do not affect spin relaxation[10,36,37]. This suggests that while we are decreasing the concentration of charge scatterers in our devices via the encapsulation process, we also unintentionally decrease the source of spin scattering in the device. For instance, this source could be magnetic scatterers which have been recently predicted to increase the spin flipping process without having a significant influence on charge transport[27,28]. In order to check the role of the magnetic impurities on spin relaxation, we now investigate the carrier concentration dependence of local device resistance and spin relaxation times in both encapsulated and non-encapsulated devices.

Prior to the carrier concentration dependent spin transport measurements, we first performed charge characterization of both encapsulated and non-encapsulated bilayer graphene junctions at room temperature. Our devices are weakly electron doped, possibly due to a charge transfer process from the MgO layer[11,12]. Mobilities of 3600 cm$^2$/Vs and 9500 cm$^2$/Vs are extracted at room temperature for the non-encapsulated and the encapsulated bilayer graphene, respectively by using $R = \frac{L}{We\mu\sqrt{n_{res}^2+n^2}}$. Here, $\mu$ is the mobility, $L$ and $W$ are the length and width of channel, respectively, $n$ is the charge carrier concentration and $n_{res}$ is the residual concentration [38]. We first note that these room temperature mobilities are higher than the similar devices measured on SiO$_2$ (See ref. 9,11 and Supplementary Figure S2), and consistent with previous works with spin-valve devices on BN where a final annealing step is not applied to keep the integrity of the spin polarized contacts[8,23,24]. It is also important to note that these mobilities are mainly limited by residues formed during the electrode fabrication process. However, the residual concentration is significantly reduced in the encapsulated junction ($n_{res}$ = 5.8 x 10$^{11}$ cm$^{-2}$) compared to the non-encapsulated one ($n_{res}$ = 1.2 x 10$^{12}$ cm$^{-2}$) since the top BN strip in the encapsulated device protects a large fraction of the junction against polymer contamination.

Figures 3a &3b show the carrier dependence of conductivity and spin relaxation time in the non-encapsulated and encapsulated junctions, respectively. The strong distortion of the conductance in the hole region is due to the metal-graphene contacts and it is common for spin valve devices[12,39]. Therefore we limit our analysis mainly to the electron region. We first discuss the low carrier concentration regime ($n$ < 4 x 10$^{12}$ cm$^{-2}$), where most of the earlier spin transport measurements were performed. Here, spin relaxation time in both the encapsulated and non-encapsulated devices has a monotonic dependence on carrier concentration. In the nonencapsulated junction, it decreases from 250 ps to 90 ps as electron carrier concentration is increased (Figure 3a). Similarly, a decrease from 330 ps to 260 ps is observed in the encapsulated junction as carrier concentration is increased in this low carrier concentration regime (Figure 3b). Such scaling is consistently observed in all devices measured both at room temperature and 2.4 K in this study (Supplementary Figure S3, S7 and S8) and it is similar to what had previously been observed at low temperatures in lower mobility bilayer graphene spin valves on SiO$_2$ substrate[9,11]. Next, we discuss the spin relaxation times at high carrier concentration regime ($n$ > 4 x 10$^{12}$ cm$^{-2}$). Around ~ 4

x $10^{12}$ cm$^{-2}$, we observe an upturn in the carrier concentration dependence of spin relaxation time in both nonencapsulated and encapsulated junctions (Colored region in Figures 3a & 3b). Remarkably there is no corresponding signature in the charge transport at this carrier concentration regime. This previously unobserved upturn in the spin lifetime brings to question what the dominant relaxation mechanism in bilayer graphene is. The earlier experiments in bilayer graphene on SiO$_2$ substrate have attributed the inverse scaling between spin and momentum relaxation times near ~ 2 x $10^{12}$ cm$^{-2}$ to the presence of Dyakonov-Perel spin scattering mechanism[11]. This low density is below the regime where the upturn in spin relaxation time is observed and therefore the scaling at this point can lead to different conclusions. We first note that neither encapsulated nor nonencapsulated devices show an inverse relation between momentum and spin relaxation times at low carrier concentration regime as shown in Figure 3c & 3d. We also observe a clear increase in spin relaxation time as the mobility of the device increases (Figure 2d). All these indicate that Dyakonov-Perel spin relaxation is not the dominant mechanism in our devices. However, at higher carrier concentration, spin and momentum relaxation times appear to change their dependence and scale inversely. Such dependence at both low and high carrier concentration regimes seem to agree with the recent theoretical work where spin relaxation in graphene is proposed to be dominated by resonant scatterers due to the presence of low concentration of magnetic impurities, such as polymer residues[27]. In that theory, the opposite carrier concentration dependence of spin relaxation in bilayer and single layer graphene was attributed to the different scales of energy fluctuations in these two systems due to their different densities of state. Away from the puddle regime, the sudden increase in spin relaxation time at ~ 4 – 5 x $10^{12}$ cm$^{-2}$ is expected for bilayer graphene due to the scattering from these resonant magnetic scatterers with no corresponding signature in the charge transport. We remark that, although our observations serve as an indication for the picture of resonant scattering mentioned above, there is no general agreement on the exact spin relaxation mechanism in bilayer graphene and further work is required to elucidate its complex nature.

Now, we turn our attention to the transport across p-n junctions in Sample B. The combination of local $V_{TG}$ and global $V_{BG}$ electrostatic gating allows us to locally control the charge density and carrier type in the channel and create bipolar junctions in between injector and detector electrodes[40]. For this, $V_{TG}$ was kept fixed at 0 V, -2.5 V and 2.5 V, while charge and spin transport measurements were performed as a function of $V_{BG}$. We first discuss the $V_{TG}$ = 0 V case. Similar to sample A, the device is slightly electron doped (Figure 4). The extracted electronic mobility at room temperature is 9000 cm$^2$/Vs. Room temperature spin relaxation time and length at $V_{BG}$ = 0 V are 400 ps and 5.75 μm respectively. Again, the spin relaxation time decreases as carrier concentration is increased. We observe clear pp'p and npn bipolar junctions while $V_{BG}$ is swept at fixed $V_{TG}$ = 2.5 V and pp'p, pnp and nn'n bipolar junctions at $V_{TG}$ = -2.5 V. Conversely, the carrier concentration dependence of spin relaxation time has very similar behavior as the local device resistance, consistent with the response in the low carrier concentration regime discussed above, considering an effective averaging of the channel properties. At present we do not find strong evidence of significant spin scattering at the pn barriers.

Finally, we study the charge and spin transport at low temperature (2 K). The presence of an on-site energy difference between the bottom and top layer induces a band gap in bilayer graphene[41]. This can be achieved with the application of a vertical electric field through the bilayer graphene plane. While infrared spectroscopy measurements yield large band gaps up to 250 meV[42], the strong role played by disorder limits the expected strong suppression of charge transport measurements and results in only a few meV of effective transport gap[43]. Figure 5a shows the 2D plot of local resistance as a function of $V_{BG}$ and $V_{TG}$. Near the charge neutrality point, the device resistance increases ~ 5-fold as the total displacement electric field is increased. This is a direct indication of the induced transport gap[41]. The extracted electronic mobility is ~ 24000 cm$^2$/Vs at 2 K. Next, we study spin transport near the charge neutrality point as the displacement field (*D*) is increased in our dual-gated device. Figure 5c&d show spin transport and spin precession measurements as *D* is increased. At *D* ~ 0, we calculate a spin relaxation length of ~ 9.5 μm. At very small *D*, we still observe a clear spin signal. The extracted spin relaxation times and lengths are in the order of ~ 500 ps and ~ 10 μm respectively, similar to the $D \sim 0$ case. This implies that no additional scattering happens at small gaps. However as $D$ increases, we observe strong suppression of spin signal and eventually the signal becomes comparable to the noise level (~ 20 mΩ) and undetectable. This prevents us from extracting the spin relaxation times at large $D$ values. Figure 5b shows the sharp decrease in the spin signal as the local resistance at the charge neutrality point increases. While a $\Delta R_{NL} \, \alpha \, R^{-1}$ dependence is expected for transparent (pin-hole) contacts in spin valve devices[33], the observation of $\Delta R_{NL} \, \alpha \, R^{-3.04}$ dependence suggests the presence of an additional mechanism driving the decrease in nonlocal signal. This resulted in a modulation of more than one order of magnitude purely via electrostatic gating (Figure 5b). We attribute this dramatic response to the formation of a transport gap in the (encapsulated) dual-gated regions. The highly resistive state in those regions would effectively block the diffusion towards the nonlocal detector of the electronic spins, which will tend to remain within the (non-encapsulated) regions where the magnetic contacts are located. This scenario leads us to an interesting case of the spin conductivity mismatch problem[44], where the regions with large and low spin conductance are both contained within the spin transport channel. We note that the spin signal at room temperature is modulated only by ~ 50 %, confirming the interpretation of a strong modulation only when a sizable transport gap is present. (Supplementary Figure S6).

## CONCLUSION

Here, we report spin transport in high mobility bilayer graphene spin valve devices encapsulated by BN. The effect of transport gap and pn junctions on spin transport is discussed. Our comparative study suggests that substrate and contacts are not the key limiting factors for spin relaxations but rather it pinpoints the role of polymer residues or ambient adsorbates. We achieve spin relaxation lengths of ~ 6 μm and ~ 10 μm at RT and 2 K, respectively. The carrier concentration dependence of spin relaxation time seems to be in good agreement with the resonant scattering-based spin scattering theory. Our flexible device architecture offers the prospect of further enhancement of the extracted spin transport parameters, by allowing optimization of the length of both the encapsulated and contact regions.

## CONFLICT OF INTEREST

The authors declare no conflict of interest.

## ACKNOWLEDGEMENTS

B. Ö. would like to acknowledge support by the National Research Foundation, Prime Minister's Office, the Singapore under its Competitive Research Programme (CRP Award No. NRF-CRP9-2011-3), the Singapore National Research Foundation Fellowship award (RF2008-07), and the SMF-NUS Research Horizons Award 2009-Phase II. I. J. V. M. acknowledges support by The Netherlands Organisation for Scientific Research (NWO). We thank S. Roche, J. Fabian, T. Taychatanapat, Y. Yeo, I. Yudhistira and S. Natarajan for their help and usefel discussions.

**FIGURE CAPTIONS**

**Figure 1** Device Characterization. **(a)** Schematics of the device preparation. **(b)** Bilayer graphene is first transferred on top of a BN layer and this step is followed by etching of graphene into a strip. **(c)** Next, a second BN layer is exfoliated onto a standard wafer and patterned into strips. **(d)** Then, the second transfer is completed by encapsulating the bilayer graphene on BN with the pre-patterned BN. The image is captured after the spin coating with PMMA to enhance the contrast of graphene. **(e)** Finally, ferromagnetic tunnel contacts are formed under ultra high vacuum conditions.

**Figure 2** Room temperature electronic spin transport measurements at $V_{BG} = V_{TG} = 0$ V. **(a-b)** Spin transport and spin precession measurements in the bilayer graphene device fabricated on the BN substrate. **(b-c)** Spin transport and spin precession measurements in the encapsulated bilayer graphene device. These two devices are adjacent and share the same bilayer graphene and BN substrate. **(d)** The mobility dependence of the spin relaxation time.

**Figure 3** Gate-dependent electronic spin transport measurements. **(a-b)** The carrier concentration dependence of the spin relaxation times for bilayer graphene- based nonencapsulated and encapsulated spin valves. The nonencapsulated device is measured at 2.4 K. The encapsulated device is measured at RT by varying $V_{TG}$, while $V_{BG}$ is fixed at 25 V. The dotted lines in the bottom panels are a guide to the eye, showing the expected spin relaxation times without involving the effect of the magnetic resonance. **(c-d)** Comparison between spin and momentum scattering times for the devices shown in Figure 3 a&b. Cyan color points represent the high charge carrier range.

**Figure 4** Bipolar electronic spin transport measurements. $V_{BG}$ dependence of the local device resistance and spin relaxation time measured at the fixed $V_{TG}$ values.

**Figure 5** Electronic spin transport measurements in the presence of a transport gap at 2 K**. (a)** 2D color plot of local device resistance as a function of $V_{TG}$ and $V_{BG}$. Device resistance at Dirac point shows 5-fold increase due to induced band gap. **(b)** Spin signal diminishes very rapidly as the transport gap is increased. Black line shows the fitting line. **(c&d)** Spin transport and spin precession measurements at various displacement fields. The measurement points are highlighted in the 2D color plot with their corresponding colors.

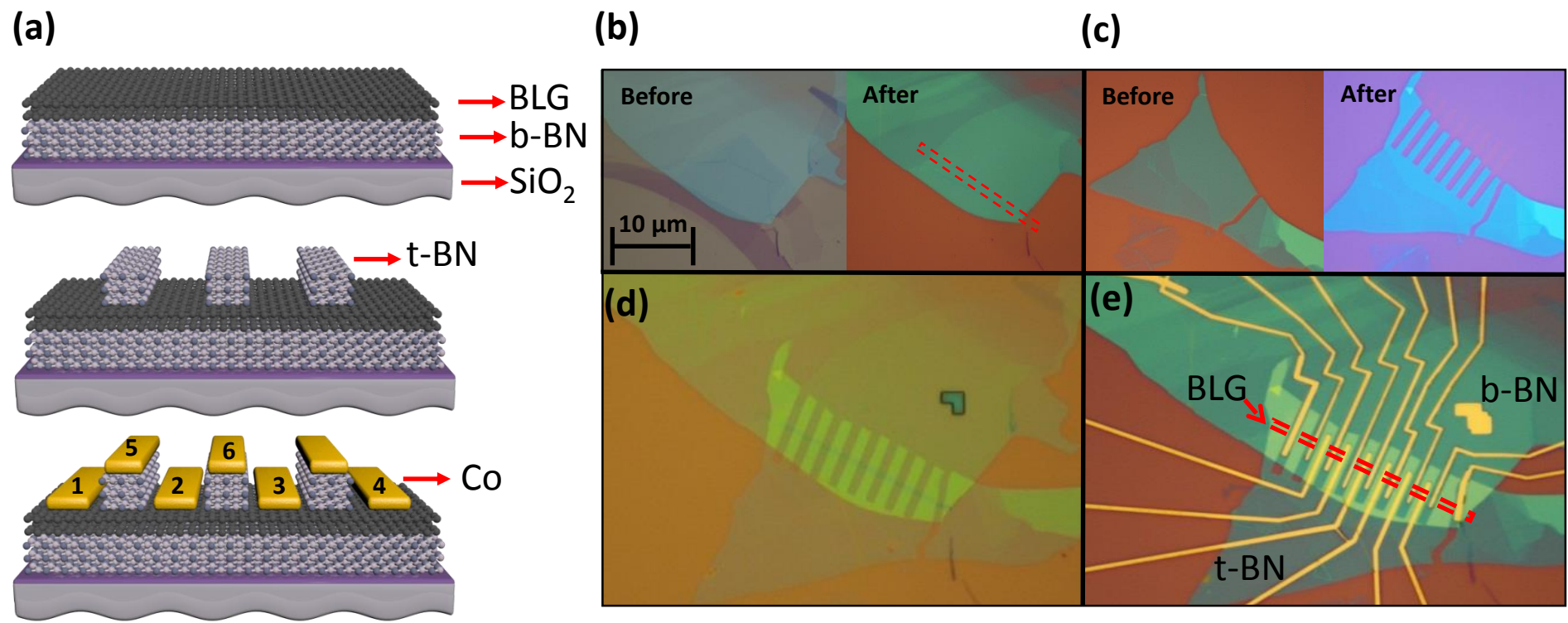

Figure 1

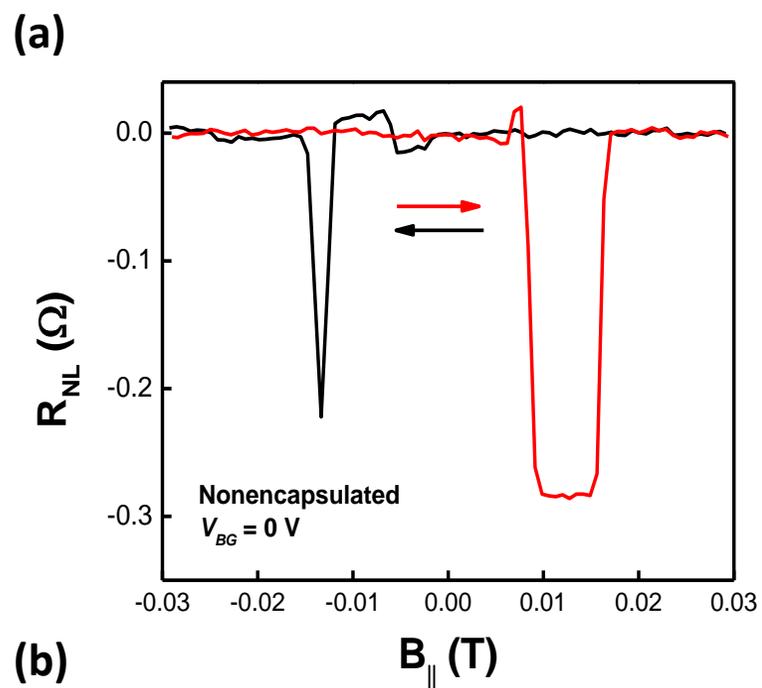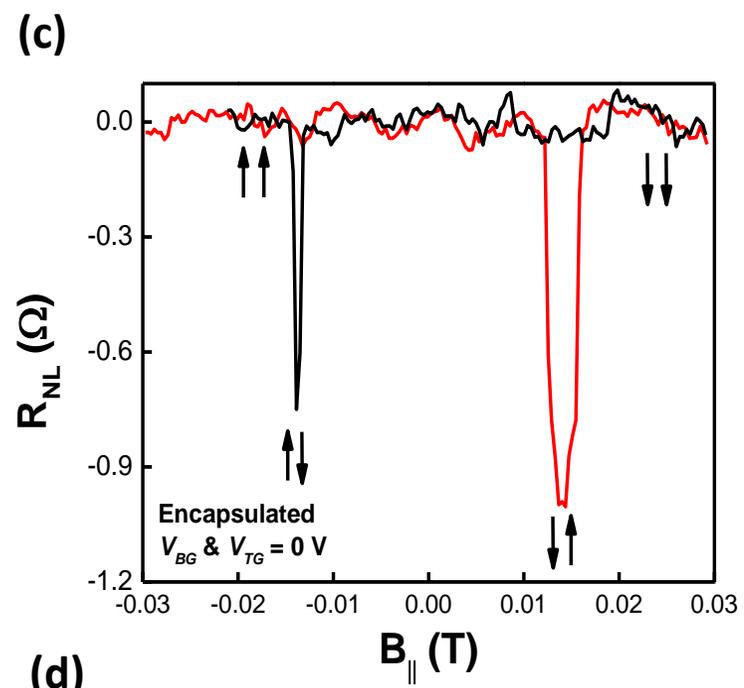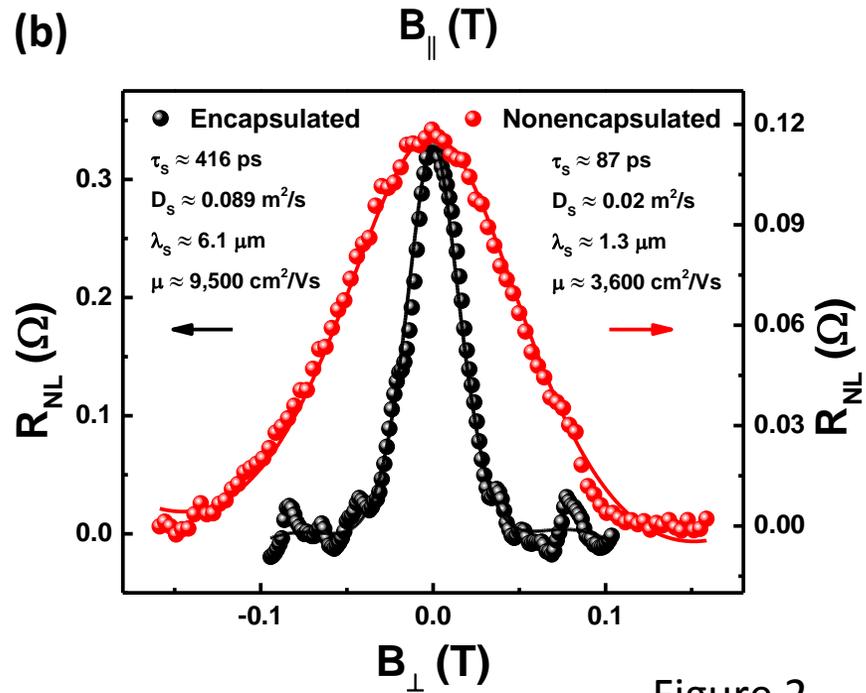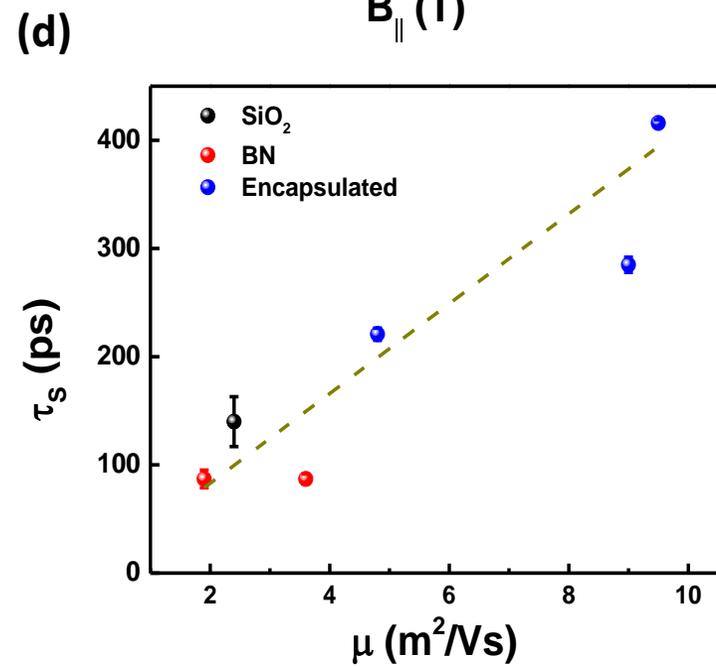

Figure 2

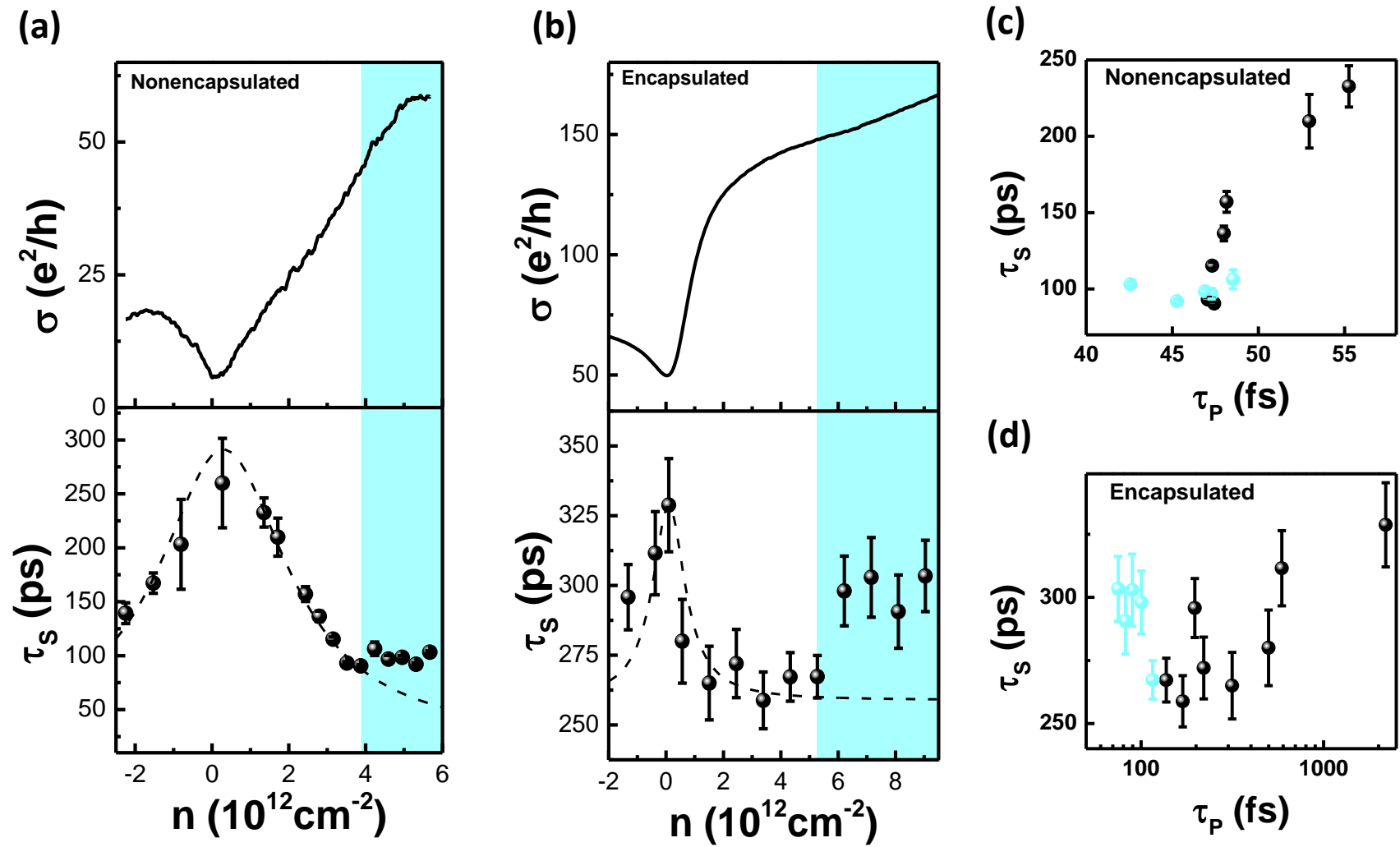

Figure 3

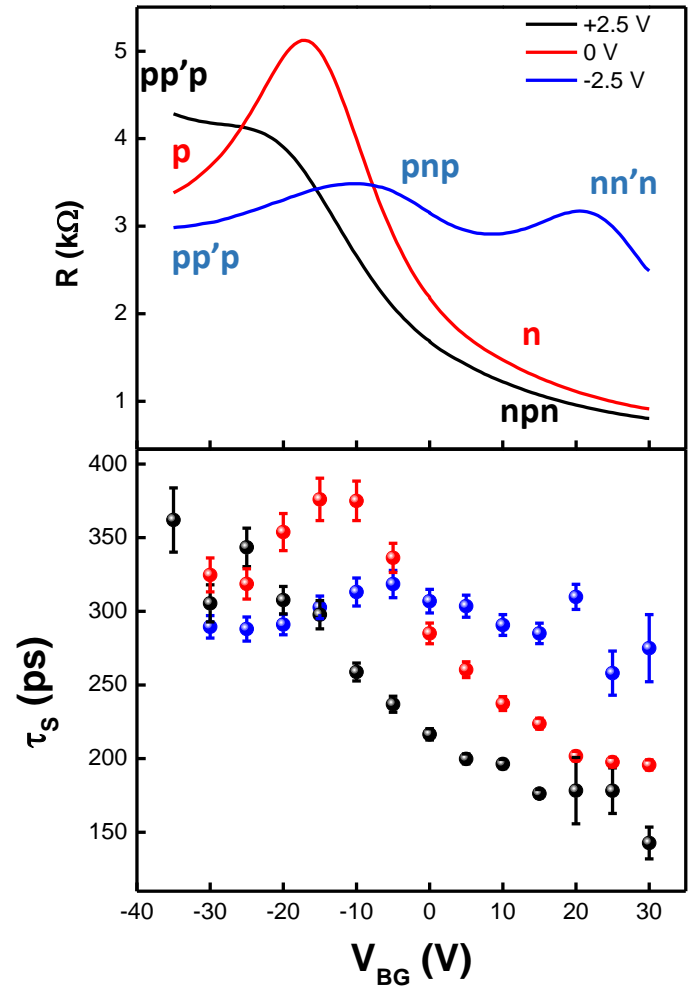

Figure 4

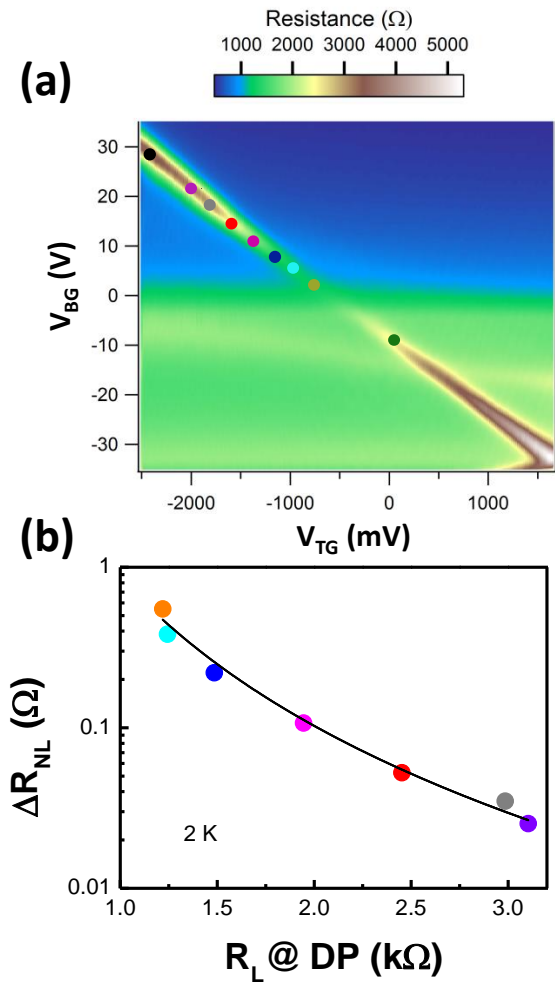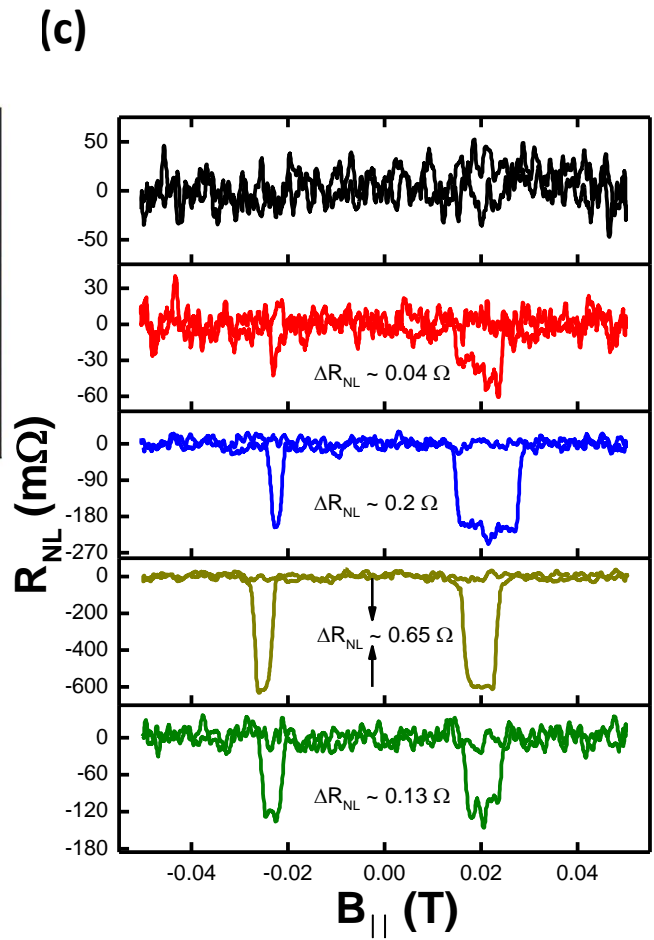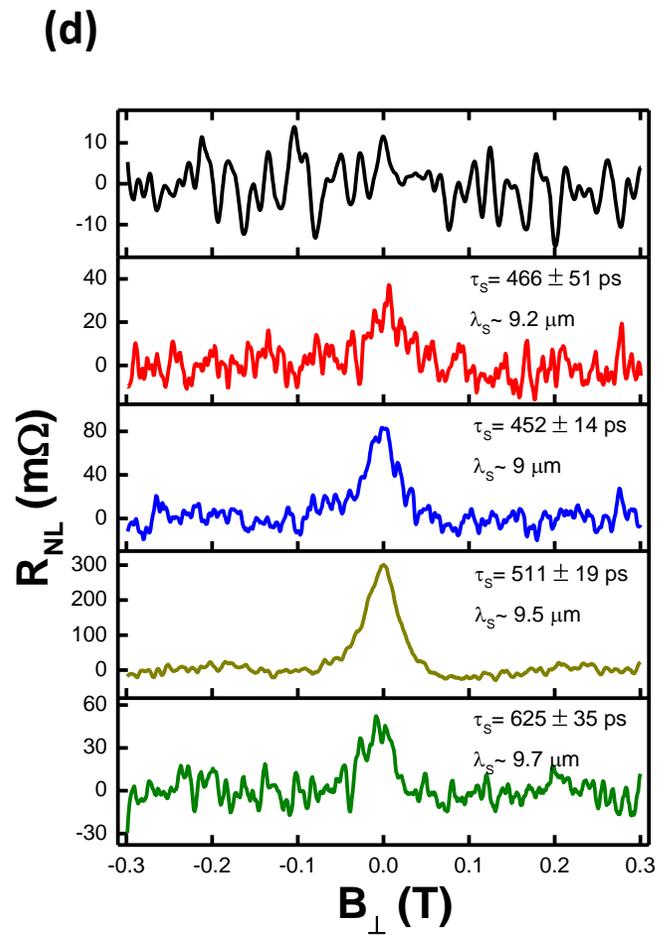

Figure 5

# Contents

# 1. Fabrication of electrodes.

A thin layer of MgO is evaporated on top of graphene, followed by ferromagnetic cobalt electrodes, in order to combat the conductivity mismatch problem. To do this, we first keep the device in ultra high vacuum (UHV) conditions for ~ 24 hours for outgassing. This is followed by annealing of the device at 100 °C for 1 hour under UHV conditions in order to minimize the concentration of PMMA residues in the patterned contact areas. We note that samples annealed higher than ~ 120 °C have lift off problem due to melting of the PMMA mask. After cooling down the sample temperature to RT, ~ 2.2 nm MgO is evaporated by using electron beam evaporation. The MgO deposition rate is ~ 0.027 A/sec. Base pressure and deposition pressure are ~ $2 \times 10^{-10}$ Torr and ~ $2 \times 10^{-9}$ Torr, respectively. To ensure the uniformity of the MgO layer, the sample is post annealed at 100 °C for 1 hour. After the device is again cooled down to RT, ~ 30 nm cobalt is evaporated under a deposition pressure of ~ $5 \times 10^{-9}$ Torr with a deposition rate of ~ 1 A/sec. A capping layer of 5 nm titanium is evaporated on top of cobalt to prevent its oxidation. Figure S1 shows the optical images of Sample A before and after the deposition process. We note that this device has different junctions to study the effect of substrate and polymer residues on spin transport.

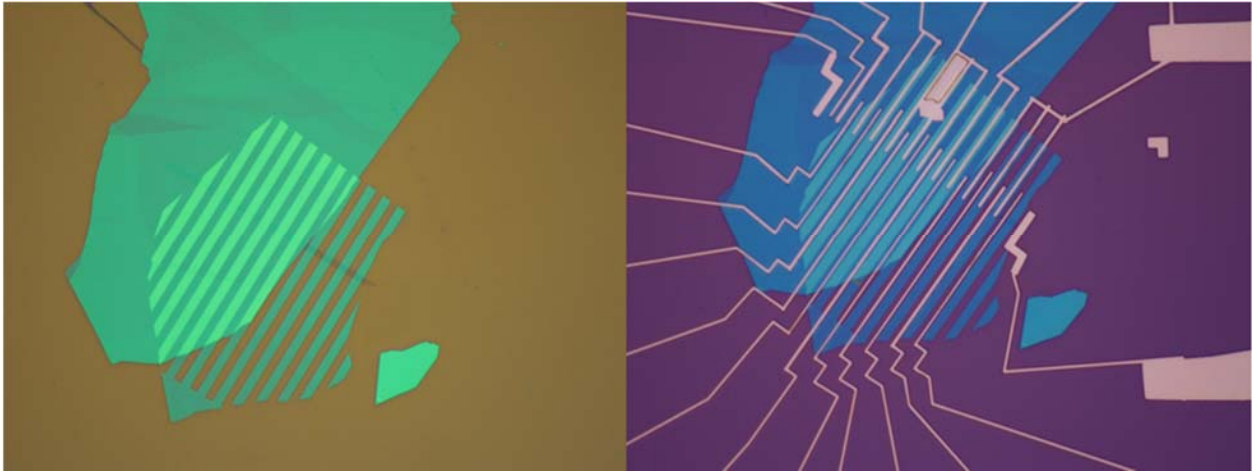

Figure S1: Optical images of Sample A before and after the deposition.

## 2. Spin transport in bilayer graphene on $SiO_2$ substrate.

Figure S2 shows the charge and spin transport measurements in bilayer graphene on $SiO_2$. Similar to the devices discussed in the manuscript, this device is also slightly electron doped. The extracted field effect mobility near $n = 1 \times 10^{12}$ cm$^{-2}$ is 2400 cm$^2$/Vs. The extracted $\tau_S$ near charge neutrality point is ~ 140 ps. This value is comparable to the device fabricated on BN substrate.

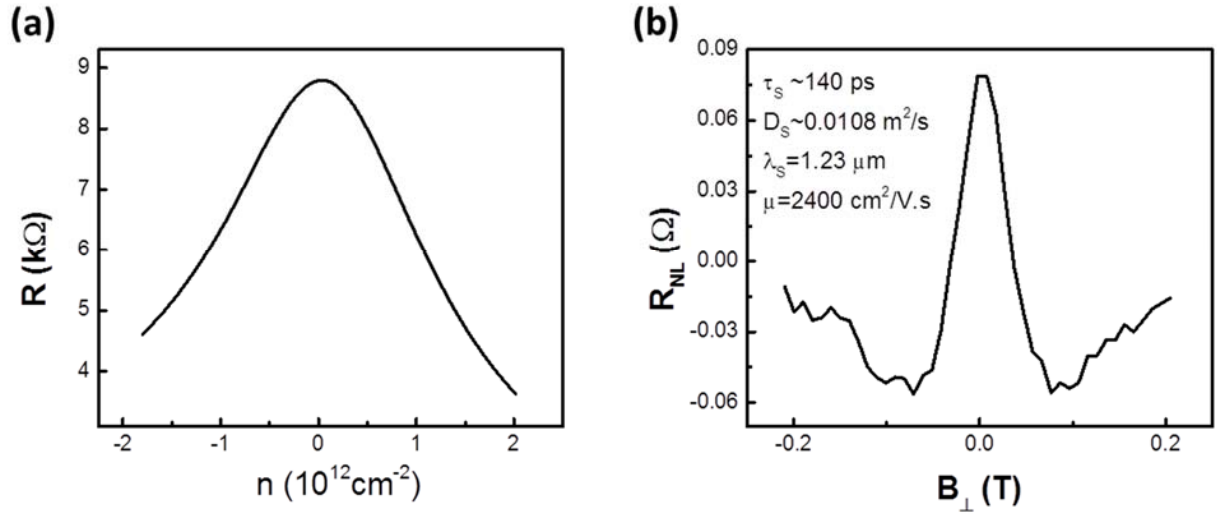

Figure S2: (a) Carrier concentration dependence of local device resistance ($R$). (b) Spin precession measurement near charge neutrality point. Measurements are performed at RT.

## 3. The carrier concentration dependence of $\tau_S$ at 2K.

Figure S3 shows the top gate voltage ($V_{TG}$) dependence of $R$ and $\tau_S$ at 2 K for Sample B. Measurements are taken at $V_{BG}$ = 4 V which is the charge neutrality point of graphene for the $V_{BG}$ sweep. Similar to RT data, the gate voltage dependence of $\tau_S$ follows the trend of $R$.

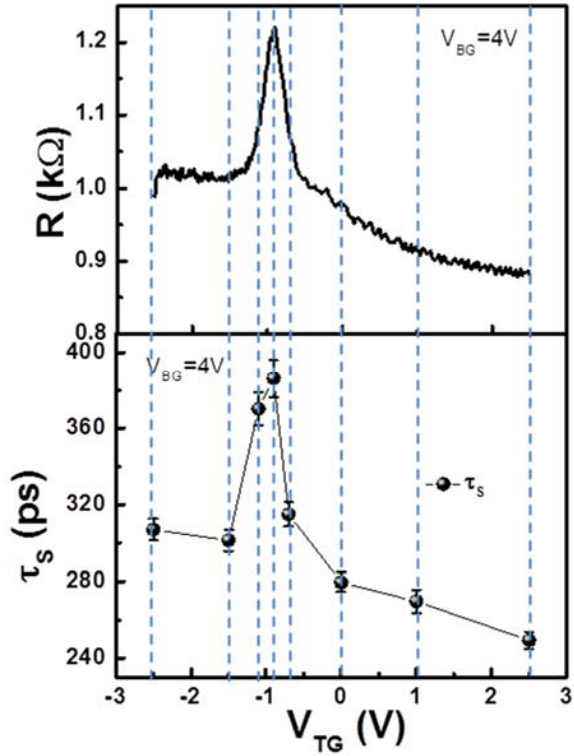

Figure S3: $V_{TG}$ dependence of $R$ and $\tau_S$.

## 4. The effect of local gating on spin transport.

In this section, we discuss the effect of locally applied top gate voltages on the spin transport. In the non-local configuration, a charge current ($I$) is sent from the injector electrode (2) to the reference electrode (1), and the electrochemical potential difference between the detector electrode (3) and the second reference electrode (4) is measured. The non-local signal ($R_{NL}$) is given by $R_{NL} = \frac{V}{I}$. The non-local configuration allows for detecting the spin dependent signal and filters the charge related spurious contributions. Figure S4b shows the RT $V_{TG}$ dependence of $R_{NL}$ and $\tau_S$ at fixed $V_{BG}$ = 0 V. Here, electrode 5 is utilized to locally modulate the charge carriers only in the charge current path. The measured $R_{NL}$ from Hanle precession measurements and the extracted $\tau_S$ seem to very weakly depending on the $V_{TG}$ and confirm that spin and charge paths are completely isolated from each other in the non-local geometry. Figure S 4c shows the $V_{TG}$ dependence of non-local resistance at $V_{BG}$ = -30 V and $B$ = 0 T, when the modulation is applied to electrode 6. While electrode 6 significantly modulates the spin signal, electrode 5 does not have any noticeable effect on spin transport.

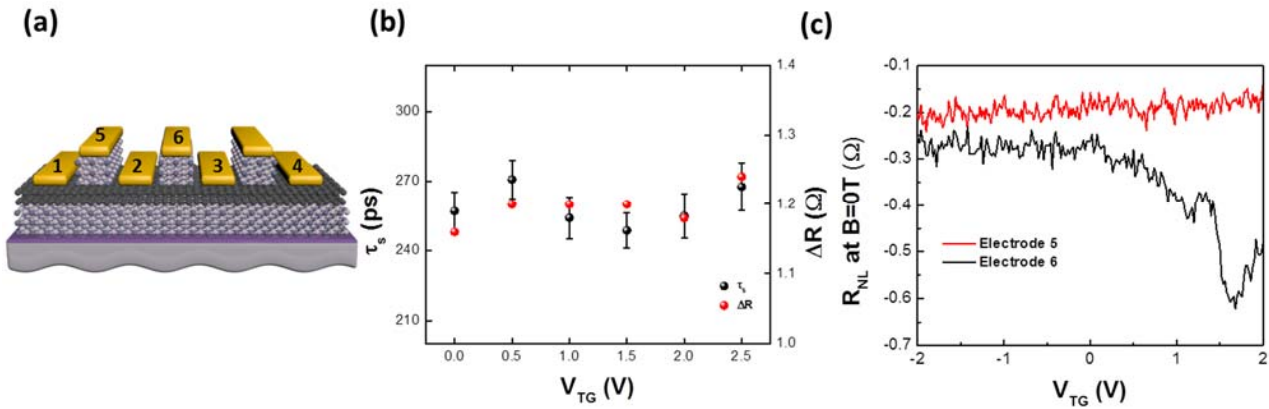

Figure S4: (a) Schematics of the dual-gated, encapsulated device. (b) $V_{TG}$ dependence of $R_{NL}$ and $\tau_S$ at $V_{BG}$ =0 V. The measurement is performed at RT in Sample B, with the modulation applied only to electrode 5. (c) $V_{TG}$ dependence of $R_{NL}$ at $B = 0$ T and $V_{BG}$ = -30 V. The measurement is performed at 2 K.

# 5. The effect of annealing on charge transport.

Annealing above 300 °C is a standard process to improve the electronic mobility of graphene by removing the fabrication-related residues. Nevertheless, such high temperature annealing degrades the ferromagnetic electrodes. Here, we show the charge transport properties of a partially encapsulated bilayer graphene spin valve device. While the encapsulated regions are protected against the fabrication residues, non-encapsulated region is being exposed to residues during the contact fabrication step. Figure S5 shows the carrier concentration dependence of the device resistance before and after an in-situ annealing at 100 °C for 12 hours. The device mobility at room temperature before and after annealing is ~ 9500 $cm^2$/Vs and 12000 $cm^2$/Vs, respectively. Such enhancement is most likely due to partially cleaning of the device from the polymer residues at the non-encapsulated regions of the junction. We note that the device mobility is ~ 48000 $cm^2$/Vs at 2 K which is the highest mobility observed in a graphene-based spin valve device up to date. Unfortunately, we could not measure any spin transport in this device after the annealing process, most possibly due to the degradation of the contact spin polarization during the annealing process.

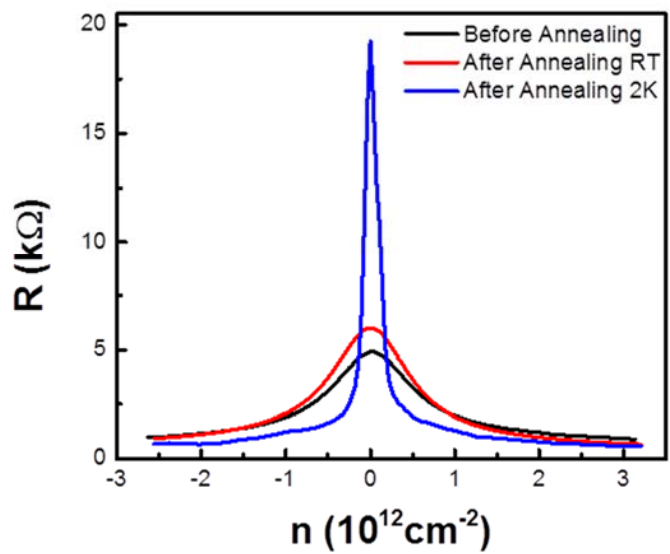

Figure S5: The carrier concentration dependence of resistance in a partially encapsulated bilayer graphene spin valve.

## 6. The carrier concentration dependence of $\Delta R_{NL}$ and the distortion on the hole conductance.

Figure S6 shows the $V_{BG}$ dependence of both local device conductivity and the amplitude of the spin signal obtained from spin-valve measurements. The relation between conductivity and spin signal indicates that our contacts are pin-hole dominated[1].

Here, it is important to mention that most of our devices show marked asymmetry on the electron and hole sides during the charge transport measurements (Figure S6). This behavior is common for graphene spin valve devices[2] and previously attributed to the graphene/contact interfaces[3]. The transfer curve in such devices consists of two contributions: graphene under the contacts with a finite n-doping and the remaining graphene channel. The n-doping to graphene results on a strong distortion of the transfer curve in the hole region[4].

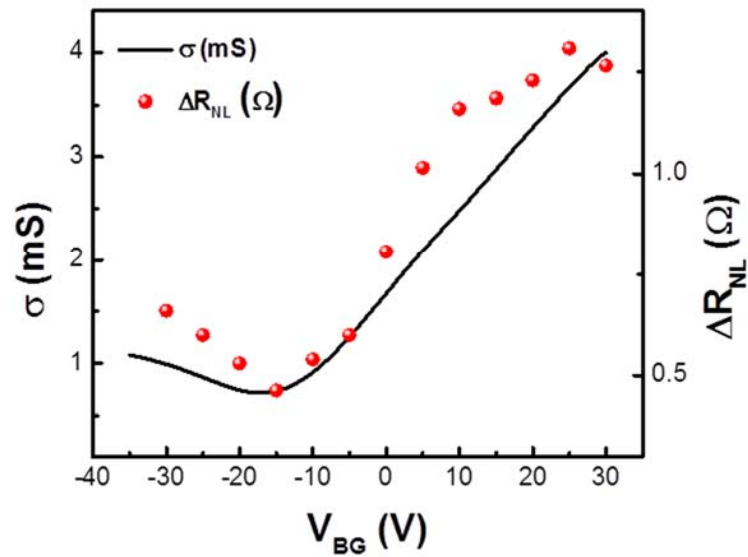

Figure S6: The carrier concentration dependence of spin signal and conductivity. The measurement is taken at RT.

# 7. The carrier concentration dependence of $R$ and $\tau_S$.

Figure S7 shows the carrier concentration dependence of both $R$ and $\tau_S$ for nonencapsulated and encapsulated device, respectively. The measurement is performed at RT. At such small carrier concentration range, $\tau_S$ depend monotonically on charge carrier concentration: it reduces as carrier concentration decreases. Note that significantly higher $\tau_S$ is observed in the encapsulated device as compared to the non-encapsulated device.

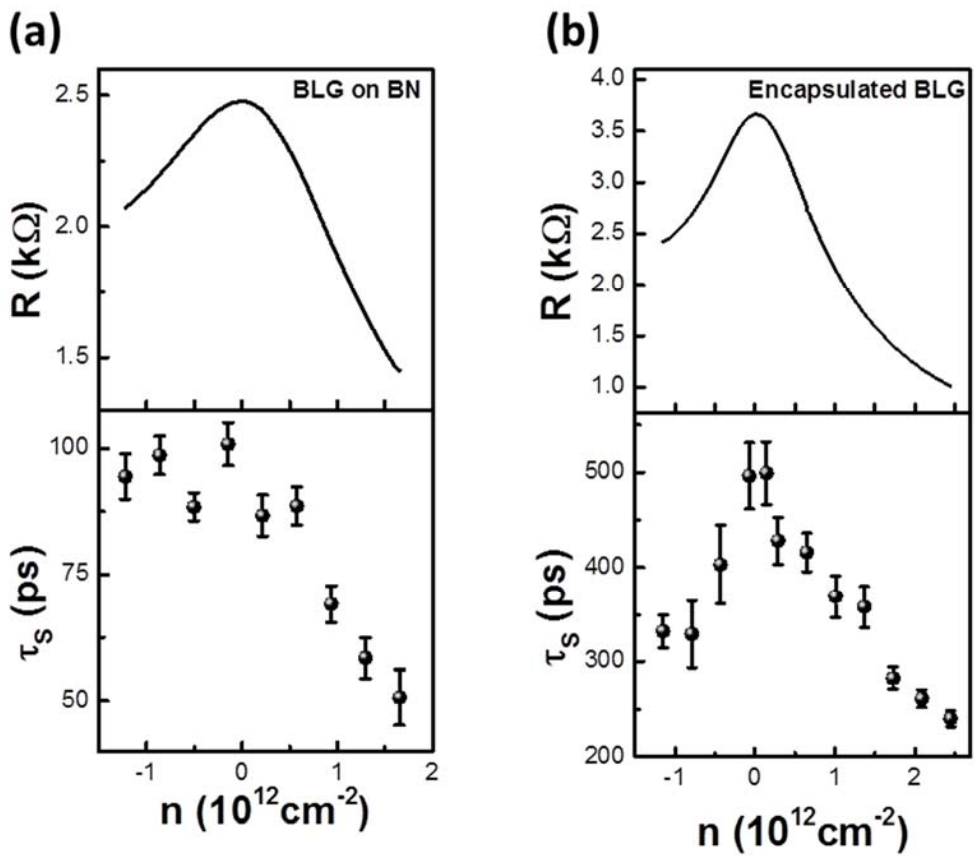

Figure S7: The carrier concentration dependence of $R$ and $\tau_S$ for non-encapsulated (a) and encapsulated (b) devices. These junctions are from the same device and fabricated adjacently.

## 8. The temperature dependence of $\tau_S$ near Dirac Point.

Figure S8 shows the temperature dependence of $\tau_S$ near Dirac Point measured in a bilayer graphene spin valve fabricated on $SiO_2$ substrate. $\tau_S$ depends on the temperature very weakly. This is consistent with the previous reports[5,6]

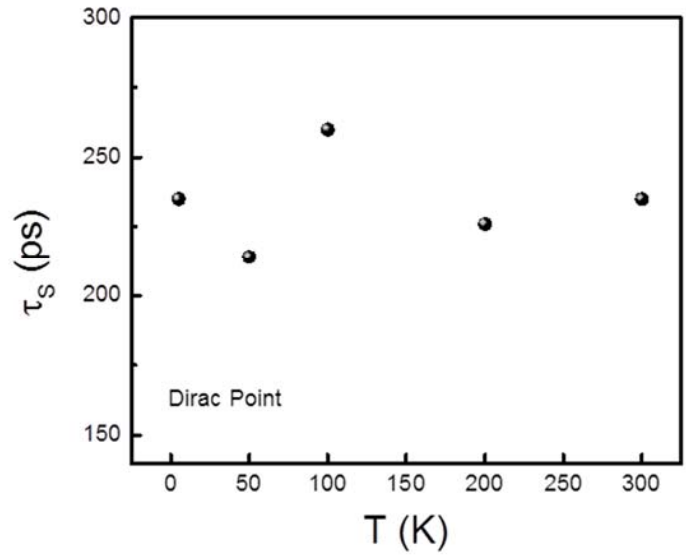

Figure S8: The temperature dependence of $\tau_S$ near Dirac Point.